\documentclass[fleqn,usenatbib]{mnras}

\usepackage{newtxtext,newtxmath}

\usepackage[T1]{fontenc}

\DeclareRobustCommand{\VAN}[3]{#2}
\let\VANthebibliography\thebibliography
\def\thebibliography{\DeclareRobustCommand{\VAN}[3]{##3}\VANthebibliography}

\usepackage{graphicx}	
\usepackage{amsmath}	

\title[Hyperbolic conduction in SPH]{Hyperbolic Conduction: A Fast, Physical Conduction Model Implemented in Smoothed Particle Hydrodynamics}

\author[Owens \& Wadsley]{
N.A. Owens,$^{1}$
J. Wadsley,$^{1}$
\\
$^{1}$Department of Physics and Astronomy, McMaster University, Hamilton, Ontario L8S 4M1, Canada 
}
\date{Accepted XXX. Received YYY; in original form ZZZ}

\pubyear{2023}

\begin{document}
\label{firstpage}
\pagerange{\pageref{firstpage}--\pageref{lastpage}}
\maketitle

\begin{abstract}


We present the first implementation of hyperbolic thermal conduction in smoothed particle hydrodynamics (SPH).
Hyperbolic conduction is a physically-motivated alternative to traditional, parabolic conduction.
It incorporates a relaxation time, which ensures that heat propagates no faster than a physical signal speed. 
This allows for larger, Courant-like, time steps for explicit schemes. Numerical solutions of the hyperbolic conduction equations require added dissipation to remain stable at discontinuities and we present a novel scheme for this.  
Test cases include a simple step, the Sod shock tube, the Sedov-Taylor blast, and a super bubble.  
We demonstrate how longer relaxation times limit conduction, recovering the purely hydrodynamical results, while  
short relaxation times converge on the parabolic conduction result.  
We demonstrate that our scheme is stable with explicit Courant-like time steps and can be orders of magnitude faster than explicit parabolic conduction, depending on the application.

\end{abstract}

\begin{keywords}
conduction -- hydrodynamics -- ISM:bubbles -- diffusion
\end{keywords}


\section{Introduction} \label{introduction}

The physics of thermal conduction and diffusion are important across a huge range of scales in astrophysics. 
This includes modelling hot gas in galaxies and clusters \citep{chandran1998}, superbubbles \citep{Weaver1977}, stellar coronae \citep{Boris2002} and cosmic rays \citep{snodin2006}. 
The traditional approach is parabolic conduction and diffusion equations, also called Fickian diffusion, named for Fick's law of diffusion \citep{fick}. 

Parabolic conduction allows information to travel instantly, which is un-physical \citep{axford1965}. 
A common ad hoc solution is to limit the flux to be less than an estimated maximum possible or saturated flux \citep{Cowie1977}. 
The saturated flux is typically set to a fraction of the heat carried if all the electrons moved in the same direction at a typical thermal electron speed relative to the gas.

With parabolic conduction,  it is assumed that the heat flux is a function of local gradients in temperature and density only. 
In practice, the flux takes a finite relaxation time to develop. 
A more physically appealing approach is to begin with the Boltzmann equations and include this relaxation time \citep{axford1965, Gombosi1993}. 
\citeauthor{Gombosi1993} show that the parabolic heat equation is a first order approximation to the Boltzmann equation.
Their second order approximation is a set of hyperbolic equations that effectively include a finite propagation speed.  
Therefore, if we switch to the hyperbolic conduction equations (sometimes referred to as Telegraph equations or Non-Fickian diffusion), we have a model that respects information propagation.  This is equivalent to having a built-in, physical limit on the flux and thus avoids the need to apply saturation limits.

The appropriate energy propagation speed depends on the system of interest. 
For neutral gas, the speed is close to the sound speed.
In the case of a plasma, electrons move faster than ions and carry most of the flux, and thus the speed should be close to that of the electrons (where the typical electron speed is roughly an order of magnitude larger than the sound speed, $c_s$) \citep{Spitzer1956}.  Magnetic fields make the propagation directionally dependent \citep{jubelgas2004}.  
However, where those same magnetic fields are tangled on small scales, the net effect is to dramatically lower the effective transport speed.  In general, we expect a transport speed related to the sound speed by a factor dependent on the physical scenario.  When the associated relaxation time is short compared to other physics, the precise factor will not strongly affect the behaviour.  In the extreme case of very short relaxation times, the behaviour becomes essentially equivalent to parabolic conduction. 

An key advantage of hyperbolic conduction is that it is significantly less computationally expensive to use in simulations. 
Explicit parabolic conduction has very stringent stability limits for its time steps, making it more expensive than other physics \citep{hanasz2003}. 
Some fluid dynamic codes circumvent the time step issue using implicit methods at the cost of an iterative solution (e.g. \citealt{Meyer2012} and \citealt{Rames2016}).
By using hyperbolic conduction, one can maintain a level of mathematical simplicity while allowing time steps similar to those of the hydro solver.  
Some authors describe hyperbolic conduction as a faster approximation to the parabolic form for numerical work \citep{Rempel2017}.  We would argue that it is both faster and more physical.

Prior numerical work has employed hyperbolic conduction in grid codes.  \citet{snodin2006} used it to model cosmic rays in the {\sc pencil} code.  
\citet{Rempel2017} and \cite{Navarro_2022} have used hyperbolic conduction to model stellar coronae.


In this paper, we will demonstrate a new implementation of hyperbolic conduction, well-suited to smoothed particle hydrodynamics, and incorporated into the {\sc gasoline2} code \citep{Wadsley_2017}.   
Hyperbolic equations typically require numerical dissipation for stability near jumps and for this we introduce a modified version of the central scheme in \citet{Kurg_Tad_2000} that takes into account the inherent diffusion of the equations. 
We demonstrate that a Courant-type time step is sufficient for explicit, stable SPH integration.   
In section three, we present results for standard test problems, including the Sod shock tube and Sedov-Taylor blast. 

Given that super bubbles are one of the processes that conduction is used to explore, we also provide this as an example. 
In the final section, we discuss potential future work.

\section{Numerical Methods}\label{Num Methods}

\subsection{Hyperbolic Conduction}\label{Hyperbolic Cond}

Parabolic conduction, is governed by the following equations,
\begin{equation}
    \mathbf{Q} = -\kappa \mathbf{\nabla}u
    \label{eq:par_cond_1}
\end{equation}
\begin{equation}
    \rho\frac{\partial u}{\partial t} = -\mathbf{\nabla}\cdot \mathbf{Q}
    \label{eq:par_cond_2}
\end{equation}
where $u$ is thermal energy per unit mass, $Q$ is the thermal flux, $\rho$ is mass density, and $\kappa$ is the thermal conductivity.  
For hyperbolic conduction, we add a time dependant component to equation~\ref{eq:par_cond_1}. This leads us to the following,
\begin{equation}
    \frac{\partial \mathbf{Q}}{\partial t} = 
    - \frac{1}{\tau}\mathbf{Q}
    -\frac{\kappa}{\tau}\mathbf{\nabla}u 
    \label{eq:hyp_cond}
\end{equation}
where $\tau$ is the relaxation time  of the system. This is traditionally known as the Telegraph equation and has been verified as a second order approximation to the Boltzmann diffusion equation \citep{Gombosi1993}.
Advective terms have been omitted here for simplicity but are included in the full implementation.
Because we use a Lagrangian method, advection is included implicitly \citep{Monaghan1992,Wadsley2004}.

In the cases where $\tau$ and $\kappa$ are kept constant, these can be rearranged into the traditional heat equation and the hyperbolic version. 
The parabolic and hyperbolic versions are as follows, 
\begin{equation}
    \frac{\partial u}{\partial t} = \frac{\kappa}{\rho}\nabla^2u
    \label{eq:par_cond_full}
\end{equation}
\begin{equation}
    \frac{\partial ^2u}{\partial t^2} + \frac{1}{\tau}\frac{\partial u}{\partial t} = \frac{\kappa}{\tau\rho}\nabla^2u.
    \label{eq:hyp_cond_full}
\end{equation}

Because it is hyperbolic, equation~\ref{eq:hyp_cond_full} has a characteristic speed that limits propagation. 
We derive this in appendix~\ref{appendix_b} as being
\begin{equation}
    c_\textrm{hyp} = \sqrt{\frac{\kappa}{\tau \rho}}.
    \label{eq:hyp_speed}
\end{equation}

To demonstrate how these equations differ, we can look at the general solution for a sample problem. 
Figure \ref{fig:frypan_exact} shows the exact solutions (as described in appendix~\ref{appendix_a}) for parabolic conduction (dashed lines) and hyperbolic conduction (solid lines) at different times.
The initial conditions are a step in temperature from 0.125 to 1 at $x=0$. 
Already, we can see that the hyperbolic transfer hits a virtual wall, limited by $c_\textrm{hyp}$, 
This virtual wall is more physical than the parabolic process. 
However, if we do wish to reproduce the parabolic case, we only need to lower our value of $\tau$ and the two systems become identical as the signal speed increases.

\begin{figure}
    \centering
    \includegraphics[width=0.48\textwidth]{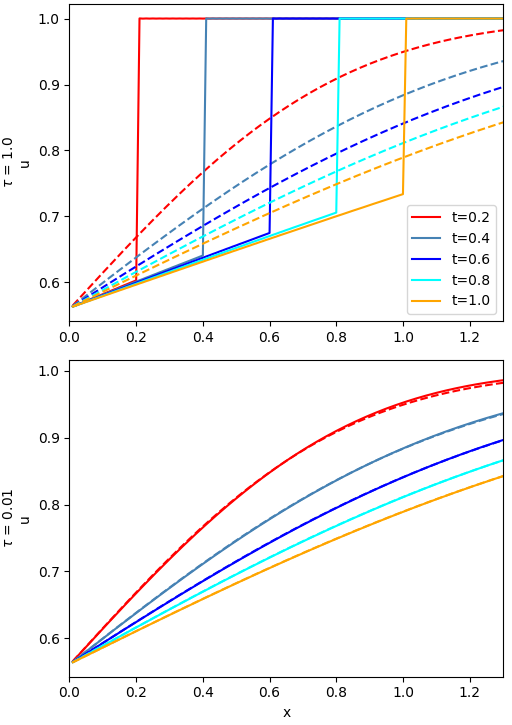}
    \caption{Analytical solutions for parabolic and hyperbolic conduction at different times. This test begins with a step in temperature from 0.125 to 1 at x=0 with a mass density of 1 and $\kappa$ = 1. Hyperbolic is shown as solid lines whereas parabolic is shown as dashed lines. The top image shows the case where $\tau$ is large enough to stop heat from propagating past a point, limited by the signal speed, $c_\textrm{hyp}$.}
    \label{fig:frypan_exact}
\end{figure}

To use these equations properly, we must determine $\kappa$ and $\tau$. 
In this first demonstration, $\kappa$ and $\tau$ are assumed to be constant. 
However, for some tests that we will perform, we are interested in the behaviour of plasma. For plasma, we use Spitzer conduction to set the $\kappa$, as described in \citet{Spitzer1956}. 
When we solve parabolic conduction explicitly, we use saturation to limit the maximum thermal conduction rates  \citep{Cowie1977}. 
Because it already limits heat transfer, there is no need to include saturation for the hyperbolic version. 
Our approach to saturation and Spitzer conductivity are described in \citet{Keller_2014}. 

Once we have $\kappa$, $\tau$ can be set using equation~\ref{eq:hyp_speed}. 
The fastest signal speed should be equivalent to the speed of the signal carrying particles (electrons) \citep{Spitzer1956}. 
We would expect this to be of on the order of the speed of sound. 
However, \citet{Cowie1977} note that their saturation equation, and therefore signal propagation, is changed depending on whether an electron's movement aligns with local magnetic fields. 
\citet{Rames2016} use a tensor conductivity to allow for anisotropic conduction along field lines.  This is beyond the scope of this work but should be feasible.
In our large scale simulations, it is commonly assumed that magnetic fields would be tangled below the limit of our resolution. 
In this case, the isotropic limit is a good representation and that the signal speed reverts to being on the order of the sound speed.
With this in mind, we can set $\tau = {\kappa}/{\rho c_s^2}$, as we would for simulations with weak or no magnetic field.

\subsection{Stable Time-Steps}

Explicit diffusion requires a time-step limit of order $0.25\,{\rho \Delta x ^2}/{\kappa}$, depending on the exact numerical representation of the diffusion term  \citep{hanasz2003}.   
The $\Delta x^2$ term is particularly troublesome. It requires that our time step goes down as the square of the resolution (e.g. particle spacing, $h$).

With hyperbolic conduction, stable timesteps become linked to signal crossing times. 
This is effectively a standard Courant-Friedrich-Lewy time step criterion \citep{Rempel2017,snodin2006}.
For hyperbolic conduction, the time-step limits is thus $\Delta t \leq \eta\, {\Delta x}/{c_\textrm{hyp}}$, where $\eta$ is a factor of order unity ($\eta \sim 0.5$). 
Keeping in mind our previous derivation, we can also express this as $\eta\, \Delta x\sqrt{{\rho\tau}/{\kappa}}$.
In cases where $c_\textrm{hyp} \approx c_s$, $\Delta t = \eta\,{\Delta x}/{c_s}$, which is satisfied via the Courant condition that is already used in {\sc gasoline2} \citep{Wadsley_2017}.

This means that our time step decreases inversely with the particle spacing, thus lengthening our time steps compared to parabolic conduction. 
The main drawback of using this method is that we now have to keep track of more variables than before and calculate more rates of change at each individual step. However, the larger time steps make up for this. 

\subsection{SPH Implementation}

In papers such as \citet{Monaghan1992}, parabolic conduction is adapted to SPH as follows, 
\begin{equation}
    \frac{du_i}{dt} = 4\sum_j\frac{\kappa_i\kappa_j}{\kappa_i+\kappa_j}\frac{m_j}{\rho_i\rho_j}\left(u_i-u_j\right)\mathbf{r}_{ij}\cdot \nabla_iW_{ij},
    \label{eq:monaghan_cond}
\end{equation}
where the right hand side is an SPH approximation for the second derivative in equation~\ref{eq:par_cond_full} and the left hand side is the comoving rate of change of the energy per unit mass, $u_i$ for particle $i$.
In this equation, $m_j$, $\rho_j$ and $\kappa_j$ are the mass, density and conductivity of particle $j$, $\mathbf{r}_{ij}$ is the vector distance between the two particles and $W_{ij}$ is the kernel function, (see \citealt{Monaghan1992}). 

For hyperbolic conduction, thermal flux can evolve separately from thermal energy and thus requires its own variable.   We adapt equations~\ref{eq:par_cond_2} and~\ref{eq:hyp_cond} separately instead of using equation~\ref{eq:hyp_cond_full}. Our SPH equations for hyperbolic conduction are shown in equations~\ref{eq:flux_sph} and~\ref{eq:thermal_sph}).

\begin{eqnarray}
    \frac{d\mathbf{Q}_{i}}{dt} &=& 
    -\frac{1}{\tau}\mathbf{Q}_i - \nonumber \\
    & & -\frac{1}{\tau}\sum_j\frac{\kappa_i\kappa_j}{\kappa_i+\kappa_j}\frac{m_j\left(\rho_i+\rho_j\right)}{\rho_i\rho_j}\left(u_i + u_j\right)\nabla_{i}W_{ij} 
    \label{eq:flux_sph}    \\
    \frac{d\,u_i}{dt} &=& \sum_j \frac{m_j}{\rho_i\rho_j}\left(\mathbf{Q}_i+\mathbf{Q}_j\right)\cdot\nabla_iW_{ij} \nonumber \\    
    & & +\, \sum_j \overline{h_{ij}}\frac{\overline{a_{ij}} m_j}{\overline{\rho_{ij}}}\left(u_i-u_j\right)\mathbf{r}_{ij}\cdot\nabla_{i}W_{ij}    
    \label{eq:thermal_sph} \\
    a_i &=& \min \left(f_1\sqrt{\frac{\kappa_i}{\rho_i \tau_i}},f_2\frac{\kappa_i}{\rho_i h_i}\right)
    \label{eq:kurg_tad_v}
\end{eqnarray}

In these equations, any variable with of the format $\overline{h_{ij}}$ is the value averaged between particles i and j.
The first term on the right hand side of equation~\ref{eq:thermal_sph} is the SPH equivalent of equation~\ref{eq:par_cond_2}. 
The second is a numerical dissipation term inspired by the dissipation term proposed by \citet{Kurg_Tad_2000}, who note that finite difference-based numerical solutions of hyperbolic equations without such terms generally experience instabilities. These take the form of ringing effects.   $f_1$ and $f_2$ are factors slightly less than unity ($f_1 \sim 0.5,f_2 \sim 0.1$)  used to optimize the amount of dissipation. 
$h_i$ is the SPH smoothing length of the particle.  If this dissipation were implemented in a grid code, $h$ could be replaced by the grid spacing.  

The added dissipation term, as outlined in \citet{Kurg_Tad_2000}, takes the form of a maximum signal propagation speed multiplied by the resolution, $\Delta x$ ($h$ in SPH), and the second derivative of thermal energy.   
A similar process was also proposed earlier by \citet{Monaghan1997} with their introduction of an artificial viscosity term into SPH.   

A twist on the usual scenario to which \citeauthor{Kurg_Tad_2000} applied their dissipation, is the fact that we are modeling a hyperbolic approximation to a diffusion equation.  Specifically, as $\tau$ gets small compared to other timescales in the system, it should closely mimic pure diffusion and finite difference-based models of diffusion equations do not need extra dissipation to remain stable.

When $\sqrt{\kappa/\rho \tau} > \kappa/\rho h$, the numerical dissipation becomes greater than the thermal diffusion associated with parabolic form.   This issue is strongly apparent in the step test of section~\ref{sec_3:step_test}.
Therefore, we limit the signal velocity to always be less than or equal to $\kappa/\rho h$.  This is equivalent to capping the signal speed when $\tau$ is small.   
We demonsrate that this is sufficient to retain stability in section~\ref{sec_3:step_test}.

\subsection{Integration Methods}

To integrate these equations, we use the leap-frog symplectic integration method described in the original {\sc gasoline} and {\sc gasoline2} papers \citep{Wadsley2004,Wadsley_2017}. 
{\sc gasoline2} uses the kick-drift-kick (KDK) form of the leapfrog. 
Velocities are kicked using calculated forces to a point half a time step ahead. 
Position is drifted to the end of the time-step using that mid-point velocity. 
Forces are then updated based on the new positions and velocities are kicked forward to the end of the time-step \citep{Wadsley2004}.   
If the rate of change of velocity depends on velocity (such as is the case for shocks or viscosity), we need a predicted velocity.  
This also applies for the flux, $Q$. 
Thus we must predict fluxes to update $Q$.  
This done using $dQ/dt$ from the previous kick.  
In the current implementation we are storing a $Q$, predicted $Q$ and a $dQ/dt$ for each particle in addition to the other variables already present in {\sc gasoline2}.

An important aspect of {\sc gasoline2} is that each particle has an individual time step which is ultimately rounded down to a power of two fraction of the largest allowed or root time step, $t_0$, so that $\Delta t = t_0/2^{n}$.  $n$ is referred to as  the rung of the particle \citep{Wadsley2004}.  Thus each full step is the product of pairwise recursive (KDK) sub-steps on higher rungs.

\section{Test Problems} \label{sample problems}

In this section, we will explore a number of tests that we can use to demonstrate the behaviour of hyperbolic conduction generally, and our SPH implementation of it.  As noted in \cite{Wadsley_2017}, lower-dimensional SPH tests do not accurately predict SPH behaviour in 3D simulations.  Thus we perform all our tests in 3D.

\subsection{Step Function}\label{sec_3:step_test}

A key test for hyperbolic conduction is a step function.  Being diffusive, numerical conduction behaves very well on smooth test problems, so we did not include such tests results here. On the other hand, the presence of an initial step can strongly excite ringing in numerical solutions of hyperbolic equations. In each test in this section, the density, $\rho$, the thermal conductivity, $\kappa$, and the relaxation time, $\tau$, were held constant.

The initial condition is of a uniform hot material with an energy per unit mass, $u=1$ placed adjacent to a cold one with $u=0.125$ with the interface at $x=0$. Hydrodynamics are not modeled.  The particles are arranged in a glass with a unit volume with $16^3$ particles replicated along the x-axis so that the particle spacing is $\sim 0.0625$.  These are thus 3D tests and are periodic with a period of $1$ in the transverse directions.  

\begin{figure}
    \centering
    \includegraphics[width=0.48\textwidth]{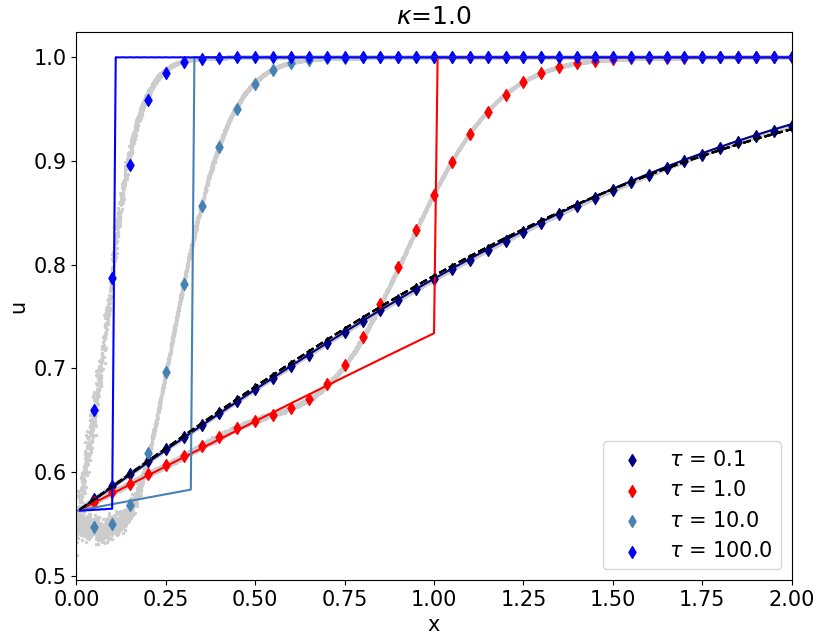}
    \caption{Hyperbolic conduction implemented in SPH. Initial conditions are a step in temperature from 0.125 to 1 at $x=0$. $\kappa = 1$, $\rho = 1$, $t = 1$, and $\tau$ varies as indicated. Solid lines are the exact solution for hyperbolic conduction. The dashed black line is the parabolic solution (this largely overlaps with the solution for $\tau = 0.1$). Diamonds are the average numerical solution solved using SPH, binned at the particle spacing. Values from individual particles are plotted in grey.}
    \label{fig:hyp_cond_1}
\end{figure}

Figure~\ref{fig:hyp_cond_1} demonstrates how the solution smoothly varies as we change the relaxation time, $\tau$, with a fixed thermal conductivity, $\kappa=1$.  
The density is kept at $\rho = 1$.
The time is set to 1 for each test so that we can see how hyperbolic conduction gradually approaches the parabolic limit as we decrease $\tau$.
In cases with a long relaxation time, the step in temperature propagates to a distance of $c_\textrm{hyp} \ t = \sqrt{1/\tau}$ (only in this case as $\rho = \kappa =1$).
In cases of small $\tau \lesssim 0.1\, t$, the parabolic and hyperbolic tests are hardly distinguishable.



\begin{figure*}
    \centering
    \includegraphics[width=\textwidth]{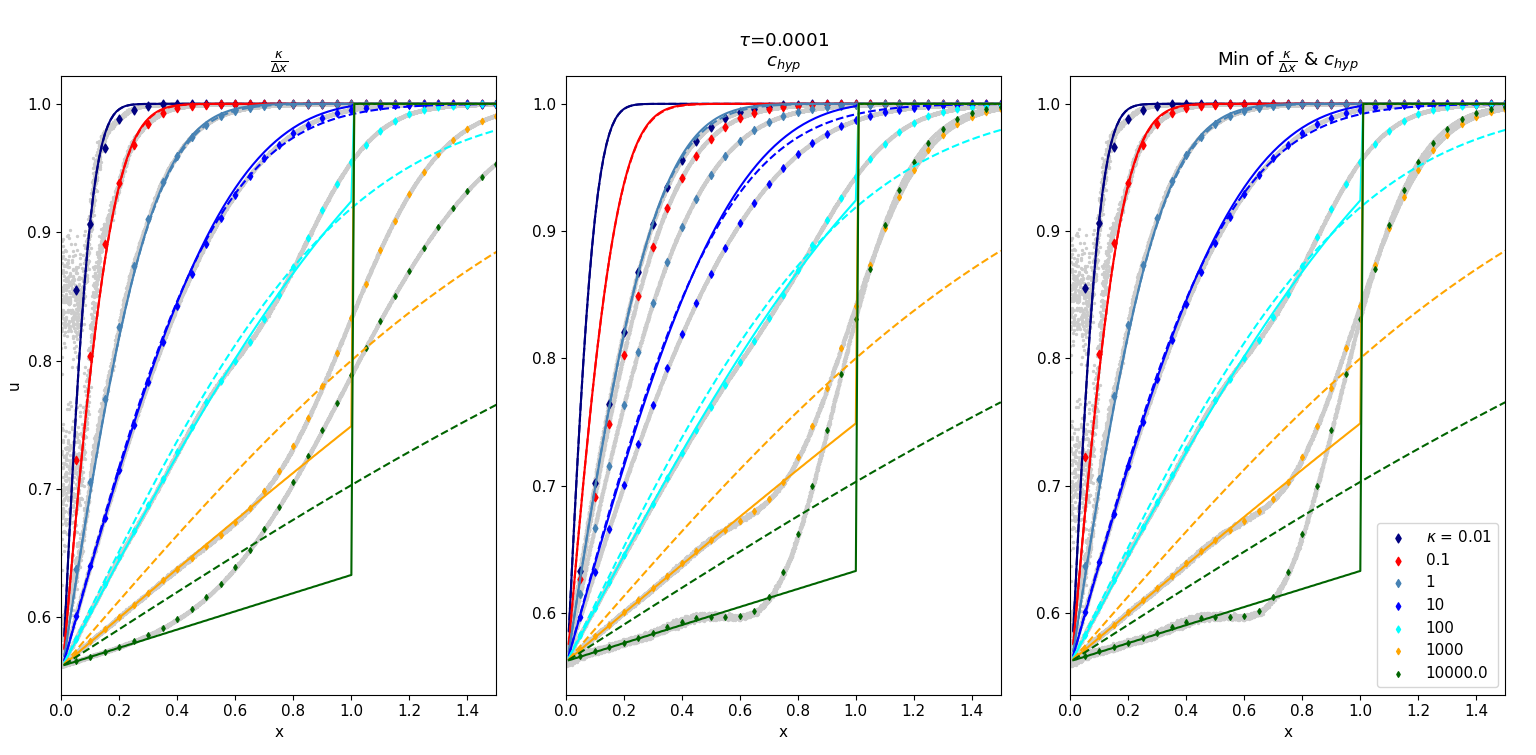}
    \caption{Step in temperature with $\tau = 0.0001$, $\rho = 0.125$, and varying $\kappa$. The left hand panel uses $0.1\kappa/\rho \Delta x$ as the characteristic speed. The middle uses $0.5\sqrt{\kappa/\rho \tau}$. The right panel takes the minimum of both. The exact solutions are represented by the solid lines. Initial conditions and time are the same as in figure \ref{fig:hyp_cond_1}. In each case, the time is set to the crossing time for a distance of 1 (e.g. $t = \sqrt{\rho\tau/\kappa} = 0.0035\sqrt{1/\kappa}$).}
    \label{fig:dispersion_demonstration}
\end{figure*}

Figure \ref{fig:dispersion_demonstration} shows a version of this test with the same step initial conditions for $u$.
Here, we compare different numerical dissipation expressions used to keep the numerical integration stable.
We also vary $\kappa$ and keep $\tau=1$ and $\rho = 0.125$ for this test.
The output time has been selected so that the step in heat always propagates a unit distance from the origin, $t = 1/c_\textrm{hyp} = 0.0035\sqrt{1/\kappa}$.

Here, we demonstrate the improvement associated with using the minimum of the two characteristic velocities for numeric dispersion expressed in equation~\ref{eq:kurg_tad_v}.

The left panel shows results using $\kappa/\rho \Delta x$ as the characteristic speed.  This is a characteristic speed for diffusion within a single resolution element.  It is too diffusive when we are far from the parabolic limit (large $\kappa$ in this case or $t \ll \tau$).
The middle panel shows a \citet{Kurg_Tad_2000} inspired scheme, using the signal speed, $c_\textrm{hyp} = \sqrt{{\kappa}/{\rho \tau}}$, as the characteristic speed.  It is too diffusive as we approach the parabolic limit (small $\kappa$ here or $t \gg \tau$).
The right panel shows results using the minimum of these two as the characteristic speed. 
Taking the minimum speed consistently recovers the exact solution. 
Although this figure only shows the solution with one value of $\tau$, tests with other values of $\tau$ produce equivalent results.

As well as the two depicted tests, we also ran tests with various other values of $\tau$ and $\kappa$ in order to calibrate the choice of the constants $f_1$ and $f_2$ in equations~\ref{eq:kurg_tad_v}.
Simulations with larger values of $\tau$ (or smaller $\kappa$) remain stable with larger time steps, and thus a shorter run-time.
The run-time is shorter than parabolic conduction, even in many of of the cases where hyperbolic conduction produces the same results as parabolic conduction.
This is the exact behaviour that we wished to achieve by introducing hyperbolic conduction.

There should be some degree of caution taken in selecting very small $\tau$ values when we wish to replicate parabolic conduction. 
We can create a system that takes longer to run than a parabolic conduction model, but this requires that $\tau$ is smaller than $\Delta x^2 \rho/\kappa$ (the diffusion time for a single element). 
From our step tests, as well as from tests in later sections, we find it is not necessary to use $\tau$ values this small for hyperbolic conduction to be a close approximation of parabolic conduction.


\subsection{Sod Shock-Tube}\label{sec_3:sod_shock}
The Sod Shock-Tube is a hydrodynamic test that has an initial jump in pressure and density but not velocity. 
We use initial conditions for the shock tube described in \citet{zamora2023}, although the shock tube is a common test that can be found in other papers including \citet{Arepo_2011}, \citet{Gottlieb1988}, and \citet{Wadsley_2017}. 
Our shock tube uses values of (1, 0, 1) for the density, velocity and pressure respectively for the left side of the shock. For the right side of the shock, we have initial conditions of (0.125, 0, 0.1). 
Our particles use the same glass layout described in section~\ref{sec_3:step_test}. 
The $\gamma$ value for both sides of the the shock is 7/5, and the thermal conductivity $\kappa$ is set to 1. 

\begin{figure*}
    \centering
    \includegraphics[width=\textwidth]{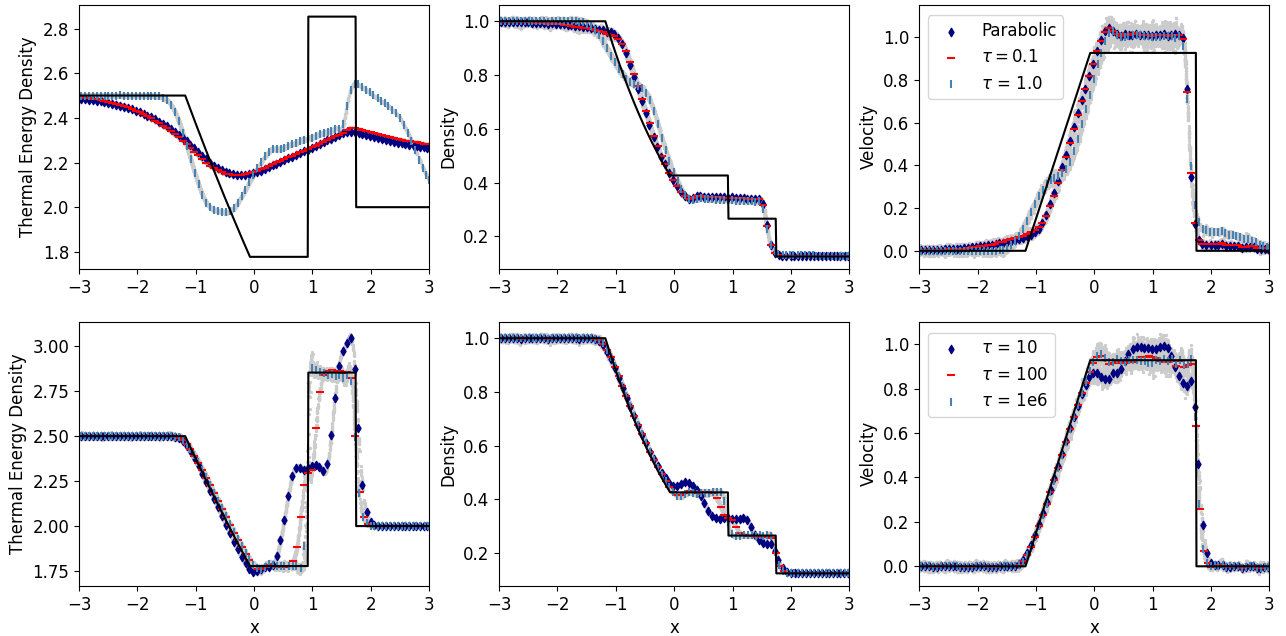}
    \caption{The conductive Sod Shock-Tube using different values of $\tau$ as well as parabolic conduction at $t=1$. Initial values of density, velocity, and pressure are (1, 0, 1) respectively on the left of $x=0$ and (0.125, 0, 0.1) on the right. The exact solution to the Riemann problem is shown in black. The top row includes cases that more closely resemble purely parabolic conduction while the bottom row shows cases closer to the non-conductive limit. 
    Values from individual particles are plotted in grey while colored particles are bins of with a physical width on the order of one particle spacing.}
    \label{fig:sod_shock}
\end{figure*}

\begin{figure*}
    \centering
    \includegraphics[width=\textwidth]{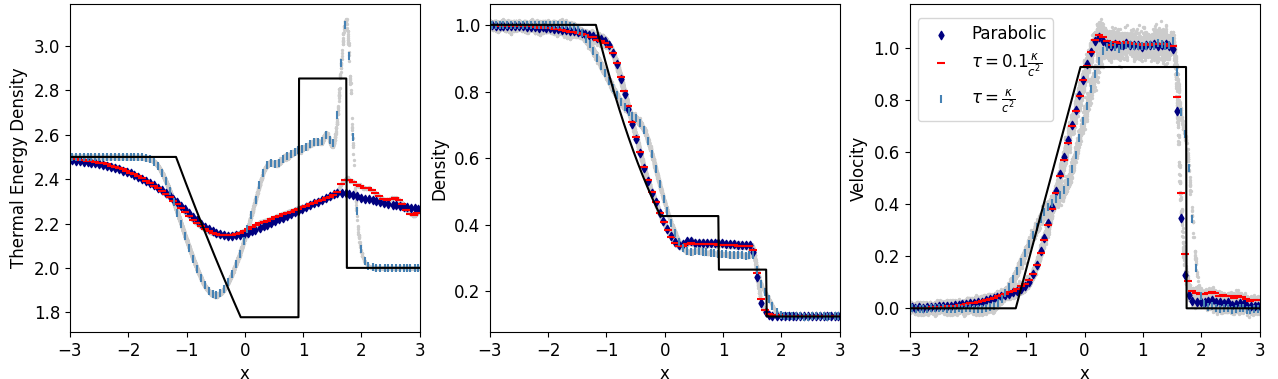}
    \caption{The Sod Shock-Tube with $\tau$ set using $\tau = 0.1\kappa/\rho c_s^2$ (red), $\tau = \kappa/\rho c_s^2$, as well as parabolic conduction. Initial conditions and time are the same as described in \ref{fig:sod_shock}. Not shown is an additional test with $\tau = 0.01\kappa/\rho c_s^2$ which almost perfectly replicates parabolic conduction.}
    \label{fig:sod_shock_c}
\end{figure*}

Our results are shown in figure \ref{fig:sod_shock}, with exact solutions to the Riemann problem shown in black for comparison. 
The test with $\tau=0.1$ is already a close match to parabolic conduction, while the cases with  $\tau \geq 100$ give similar results to a non-conductive (purely hydrodynamical) shock-tube. 
Because we would like to limit conduction to the sound speed, we also tested the shock-tube in the case where $\tau = \kappa/\rho c_s^2$. 
This is shown in figure \ref{fig:sod_shock_c}. 
In this shock-tube, the sound speed varies between roughly 0.9 and 1.1.

We see promising results right away. 
As we can see in figure \ref{fig:sod_shock}, the case where $\tau = 0.1$ mimics parabolic conduction. 
Intuitively, we would expect this because $\tau$ is much less than $t$.

For the sound-speed derived relaxation time shown in figure \ref{fig:sod_shock_c}, reducing the the coefficient on the relaxation time by a factor of 10 causes it to almost perfectly reproduce parabolic conduction. 
In general, this may be a valuable rule of thumb to use. $\tau \sim 0.1\kappa\rho/ c_s^2 \lesssim t$ nearly replicates parabolic conduction. 
There is some additional noise in the low $\tau$ case but the system remains stable.  This gives a signal speed which is about 3 times the sound speed, with correspondingly smaller numerical time steps.

In terms of computational expense, the minimum time step for parabolic conduction is 128 times smaller than the minimum time step for the $\tau = 0.1\kappa/\rho c_s^2$ test (being a power 2 due to our time stepping scheme). 
Every other hyperbolic has runs with higher time steps than this. 
This is a huge increase in the speed; it can be the difference between a test taking hours and minutes to run. 
Therefore, even if we would like to perfectly replicate parabolic conduction, a hyperbolic system with low $\tau$ remains the better option.

Finally, our replication of the non-conductive system shows the level of control hyperbolic conduction provides over the conductive properties of the system, with only small amounts of diffusion compared to the exact solutions. 
Thus by varying the relaxation time, $\tau$, we can explore disparate physical regimes (and even effectively fully suppress conduction if we wish).

\subsection{Sedov-Taylor Blast}\label{sec_3:sedov}
The Sedov-Taylor Blast is an explosion test dating back to the 1940's \citep{Taylor1950,sedov1959}. 
The test case offers a very strong shock with an analytic solution. 
Because it is a strong shock, as mentioned in \citet{Wadsley_2017}, we must select our initial conditions very carefully. 
For the ideal Sedov-Taylor Blast, for which we have an analytical solution, we would begin with a point explosion and the background temperature would be 0. 
Our test uses 64 $4.2\times10^8$ kelvin ($\sim 10^{55}$ ergs) particles placed in the centre of the box.
This box is a glass of $128^3$ particles. 
The surrounding particle energies correspond to $10^{-3}$ kelvin, a number well below the central temperatures.

\begin{figure*}
    \centering
    \includegraphics[width=\textwidth]{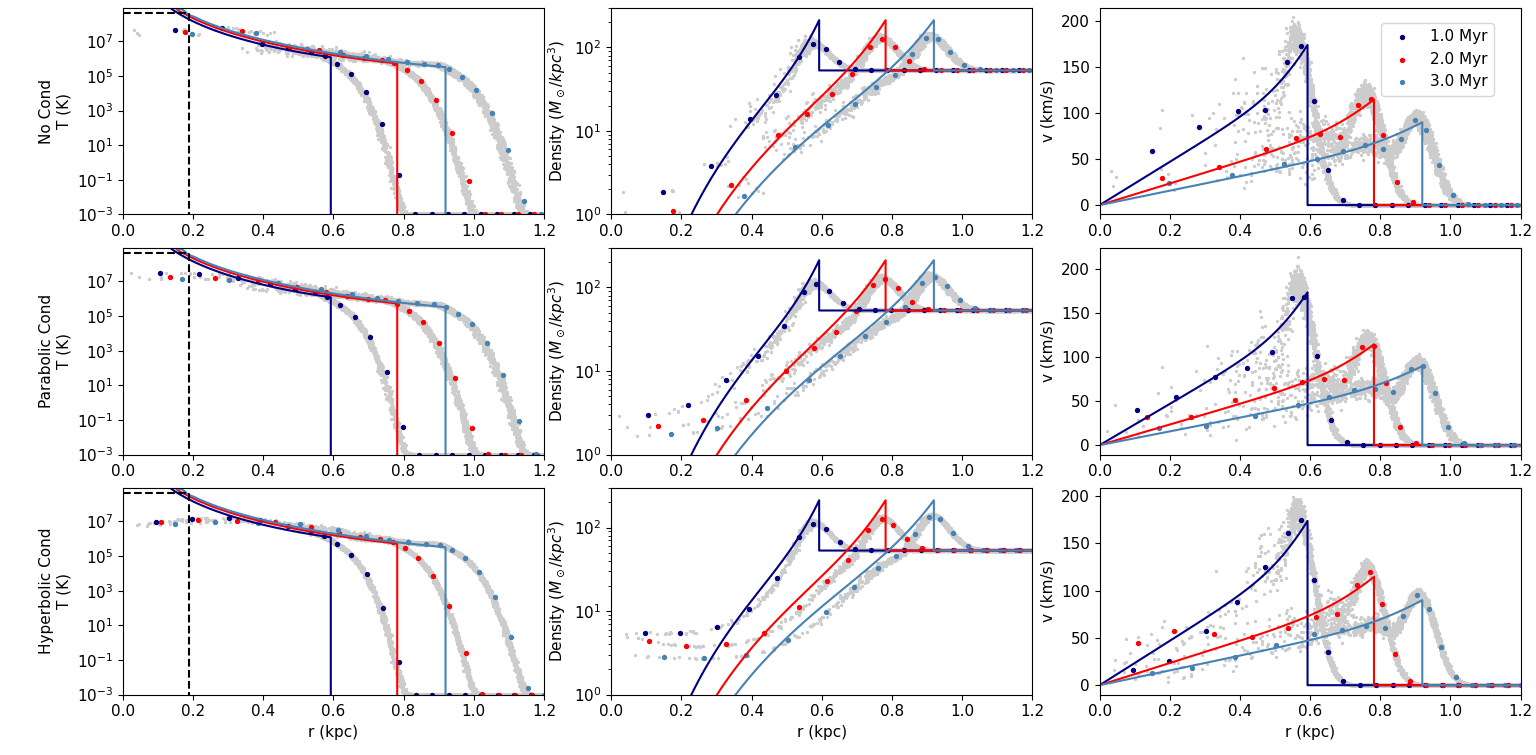}
    \caption{The Sedov-Taylor blast at 1, 2, and 3 Myr. This test includes a plasma of $4.2\times10^8$ kelvin gas in the centre of a cold box. The top row shows no conduction while the bottom two show parabolic and hyperbolic conduction respectively. Solid lines show the exact answer for a point explosion as described in \protect\citet{Taylor1950,sedov1959}. Black dashed lines show the initial distribution of temperature for our particles.}
    \label{fig:sedov}
\end{figure*}

Figure~\ref{fig:sedov} shows the resulting blasts at three times for the no conduction (top row), parabolic conduction (middle row), and hyperbolic conduction (bottom row). 
Because this test is meant to represent an explosion within a real plasma,  we use the equation for Spitzer Conduction to set $\kappa$ and sound speed to set $\tau$ \citep{Spitzer1956,Cowie1977}. 
The results near the shock-wave are similar with and without conduction.
Inside the bubble, we see a small increase in density at small radii, corresponding to slightly lower temperatures.   The amount of affected mass is very small.

Examining the individual particle timesteps, we find that the minimum time step used for the parabolic case was 16 times smaller than the hyperbolic case.  For completeness, we note that the conduction free (purely hydrodynamic test) had a minimum time step slightly larger than the conductive runs.   Overall, the computational effort for the parabolic run was much larger than the hyperbolic case.

The large temperature contrast of the Sedov-Taylor blast produces strong conduction with high characteristic velocities entering the numerical dissipation expression.  This results in more temperature diffusion in the high temperature, low density central region in the hyperbolic case compared to the parabolic case.   This lowers the temperatures slightly (as seen in the figure).

The blast appears to show little change at the shock due to conduction, which is expected as the temperatures are low enough to limit Spitzer conduction.

\subsection{Superbubble}\label{sec_3:superbubble}

\citet{Weaver1977} first showed that thermal conduction can play an important role in the formation of wind-driven bubbles around star clusters. 
Conduction allows additional hot mass to be fed into the bubble  interior, changing its temperature \citep{Weaver1977}. 
This model also applies for supernova driven bubbles.  It provides a basis for energy-based, sub-grid models for stellar feedback \citep{Keller_2014}.

\begin{figure*}
    \centering
    \includegraphics[width=\textwidth]{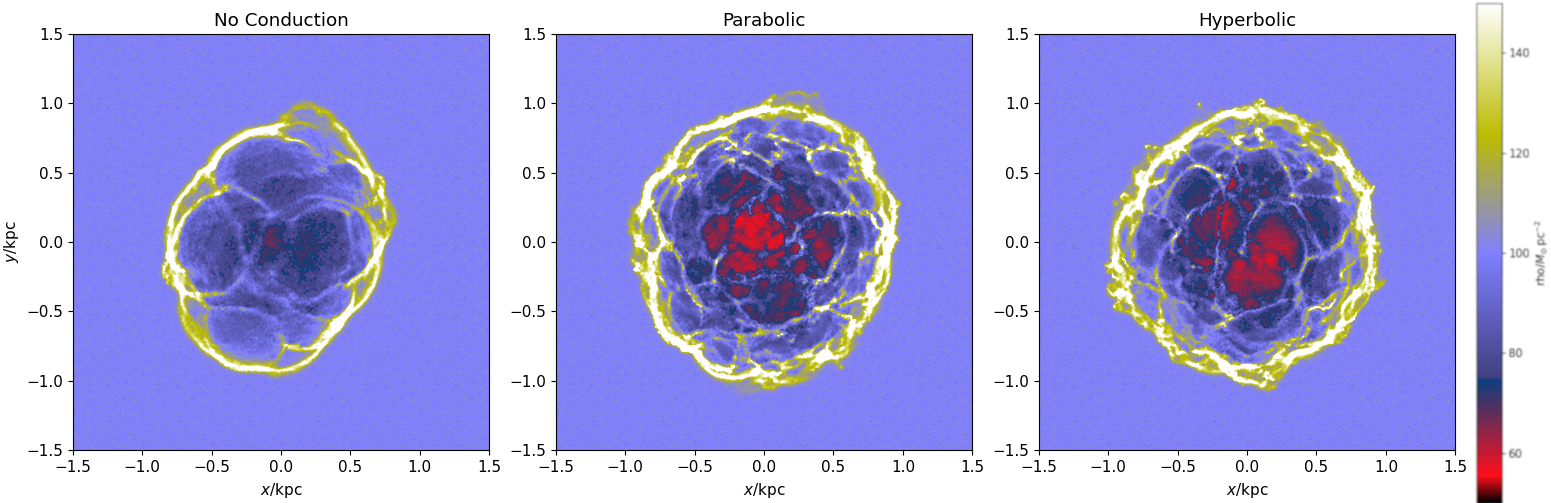}
    \caption{A super bubble test 50 Myr after the creation of the star cluster. Hot mass is continuously injected from stellar winds in a central 30,000 $M_\odot$ star cluster. The top left test is run with no conduction. The centre test uses parabolic conduction. The rightmost test uses hyperbolic conduction with $\tau = \kappa/\rho c_s^2$. The conductive bubbles grow much larger than their non-conductive counterpart and create larger instabilities.}
    \label{fig:superbubble}
\end{figure*}

Our test is shown in figure \ref{fig:superbubble}. We continuously inject hot particles into the centre of our system to simulate hot supernova ejecta from a star cluster of 30,000 solar masses.  The star cluster particle is placed in the centre of a 4x4x4 kpc box with $128^3$ particles. 
This closely follows the direct injection test demonstrated in \citet{Keller_2014}, which we aim to reproduce with hyperbolic conduction.
From left to right, the panels show the non-conductive case, parabolic conduction, and hyperbolic conduction.

There is a difference in the size of the bubbles depending on whether or not we include conduction. 
Hyperbolic conduction produces similar behaviour to parabolic conduction. 
The conductive cases have higher levels of hot mass injected back into the bubble.  The overall result is more mass inside the bubble and a larger bubble in the conductive cases, as shown in the right two panels of figure~\ref{fig:superbubble}. 
Although the projected density inside the bubble is increased in non-conductive case, analysis shows that the mass density inside the bubble is similar in all three cases.
The conducting results match those of \citet{Keller_2014}.  

We see differences in run time.
For parabolic conduction, the minimum time step is 16 times smaller than the minimum time-step for both the hyperbolic and non-conductive run which have the same minimum time step.
The increase in speed from switching to hyperbolic conduction is clear.

Our superbubble tests also demonstrate the effect of conduction on an expanding bubble. 
Vishniac instabilities are expected to grow on the surface of the bubble, just as we see in figure \ref{fig:superbubble} \citep{vishniac1983}. 
These instabilities arise from the cool shell of the superbubble interacting with the surrounding medium,  They are more developed in cases where the bubble is conductive.

\section{Discussion \& Conclusions}\label{Discussion and Conclusions}






\subsection{Future Work}

Our hyperbolic conduction implementation uses more numerical dissipation near discontinuities than the parabolic version.  For example, there is a small temperature decrease in the centre of the Sedov-Taylor blast.
Our discontinuity model (and associated parameters) were calibrated to supply enough dissipation to remain stable on tests such as the shock tube (section~\ref{sec_3:sod_shock}).  
However, with large temperatures, Spitzer conductivity can lead to large characteristic speeds and correspondingly large numerical dissipation.
Our numerical dissipation is modeled after the central scheme of \citet{Kurg_Tad_2000}, who apply it using higher order methods such as linear reconstruction.
If we were to use piecewise linear rather than piecewise constant reconstruction we can probably lower the dissipation substantially, particularly for non-zero gradients, as suggested by \cite{beck2016}.  
It would be worth exploring this as an addition to {\sc gasoline}2 in the future.

So far, we have assumed that magnetic fields play no direct role in the conductive process. 
As noted previously, unresolved, tangled magnetic fields can effectively make the diffusion isotropic.   More generally, 
if we wish to use hyperbolic conduction with magnetic fields, we should use anisotropic conduction similar to that found in \citet{Rames2016}.  This also applies when modeling cosmic ray diffusion \citep{snodin2006}.

A key driver for this work has been to facilitate exploration of the role of superbubbles as a mode of energetic stellar feedback (from winds and supernovae).  
The hot mass in the bubble and its temperature are critically dependent on conduction.   
Once we move beyond idealized superbubbles \citep{Weaver1977} to the real ISM, we need to consider a clumpy medium, magnetic fields affecting conduction, unresolved cooling and turbulent conduction.  These effects were explored in one dimension by \cite{Badry2019}.   
This is a promising potential application of our research.    
For example, \citet{Keller_2014} developed a sub-grid two-phase model based on idealized superbubbles which could benefit from an update that incorporates these non-ideal effects.

\subsection{Summary}

We have argued that hyperbolic conduction is a viable, and physically more appealing, alternative to using parabolic conduction.  We have also demonstrated that it works in practice with a robust SPH implementation.

Hyperbolic conduction has the advantage that we can limit the propagation of information to the relevant physical signal speed.  Thus changes cannot propagate instantly to any distance which is the primary unphysical aspect to the parabolic form.  A related benefit is that we do not need to contemplate applying a saturation cap to the allowed fluxes.   An advantage for simulations is that numerical stability only requires a simple Courant-type time step limit based on the signal speed.  This allows us to use larger time steps (similar to those for hydrodynamics) and therefore run our simulations faster. 

We implemented hyperbolic conduction within SPH.  Traditionally, SPH uses explicit numerical dissipation to handle discontinuities (such as artificial viscosity for jumps in the hydrodynamic equations).  We found that a standard prescription based on the hyperbolic signal speed created excess diffusion.  We demonstrated that the intrinsic diffusion in the hyperbolic conduction equations allows us to use progressively lower numerical dissipation while remaining numerically stable as the relaxation time becomes shorter and we approach the parabolic regime.  

In all cases, the time step limit remains Courant-like, $\Delta t  \lesssim \Delta x/c_\textrm{hyp}$, and our code ran stably without the need for the rather prohibitive $\Delta t \lesssim (\Delta x)^2/\kappa$
time step associated with parabolic diffusion.  We note that hyperbolic conduction brings some computational overheads, such as the need to store and evolve the vector flux quantity.

A hyperbolic conduction signal speed comparable to the sound speed generally gives results  that are qualitatively and quantitatively similar to the parabolic case.  There are good physical arguments to support the heat transport speed being similar to the sound speed.  In many test problems, there are other physical timescales that are much slower than these speeds and the precise choice of speed has limited effects.  In these cases, the sound speed is an attractive choice.
However, the precise signal speed may be higher (e.g. a plasma assuming no magnetic fields).   The user might also desire a very close match to the parabolic result, which generally requires a signal speed 3 times the sound speed or faster.   In such cases, the hyperbolic approach is still typically more computationally efficient than an explicit parabolic approach.

\section*{Acknowledgements}

The analysis was performed using the pynbody package (\url{https://github.com/pynbody/pynbody}, \citep{pynbody}). Further analysis was done using pytipsy by Ben Keller (\url{https://github.com/bwkeller/pytipsy}, \citep{Pytipsy}). The simulations were performed on the clusters hosted on sharcnet, part of Compute Canada. We greatly appreciate the contributions of these computing allocations.

\section*{Data Availability}

Data provided in this is generated using {\sc gasoline}2 by \citet{Wadsley_2017}. {\sc gasoline} is available as a public release from \url{https://gasoline-code.com/}.



\bibliographystyle{mnras}
\bibliography{biblio.bib} 

\begin{thebibliography}{}
\makeatletter
\relax
\def\mn@urlcharsother{\let\do\@makeother \do\$\do\&\do\#\do\^\do\_\do\%\do\~}
\def\mn@doi{\begingroup\mn@urlcharsother \@ifnextchar [ {\mn@doi@}
  {\mn@doi@[]}}
\def\mn@doi@[#1]#2{\def\@tempa{#1}\ifx\@tempa\@empty \href
  {http://dx.doi.org/#2} {doi:#2}\else \href {http://dx.doi.org/#2} {#1}\fi
  \endgroup}
\def\mn@eprint#1#2{\mn@eprint@#1:#2::\@nil}
\def\mn@eprint@arXiv#1{\href {http://arxiv.org/abs/#1} {{\tt arXiv:#1}}}
\def\mn@eprint@dblp#1{\href {http://dblp.uni-trier.de/rec/bibtex/#1.xml}
  {dblp:#1}}
\def\mn@eprint@#1:#2:#3:#4\@nil{\def\@tempa {#1}\def\@tempb {#2}\def\@tempc
  {#3}\ifx \@tempc \@empty \let \@tempc \@tempb \let \@tempb \@tempa \fi \ifx
  \@tempb \@empty \def\@tempb {arXiv}\fi \@ifundefined
  {mn@eprint@\@tempb}{\@tempb:\@tempc}{\expandafter \expandafter \csname
  mn@eprint@\@tempb\endcsname \expandafter{\@tempc}}}

\bibitem[\protect\citeauthoryear{{Axford}}{{Axford}}{1965}]{axford1965}
{Axford} W.~I.,  1965, \mn@doi [\planss] {10.1016/0032-0633(65)90063-2}, \href
  {https://ui.adsabs.harvard.edu/abs/1965P&SS...13.1301A} {13, 1301}

\bibitem[\protect\citeauthoryear{{Beck} et~al.,}{{Beck}
  et~al.}{2016}]{beck2016}
{Beck} A.~M.,  et~al., 2016, \mn@doi [\mnras] {10.1093/mnras/stv2443}, \href
  {https://ui.adsabs.harvard.edu/abs/2016MNRAS.455.2110B} {455, 2110}

\bibitem[\protect\citeauthoryear{{Chandran} \& {Cowley}}{{Chandran} \&
  {Cowley}}{1998}]{chandran1998}
{Chandran} B. D.~G.,  {Cowley} S.~C.,  1998, \mn@doi [\prl]
  {10.1103/PhysRevLett.80.3077}, \href
  {https://ui.adsabs.harvard.edu/abs/1998PhRvL..80.3077C} {80, 3077}

\bibitem[\protect\citeauthoryear{{Cowie} \& {McKee}}{{Cowie} \&
  {McKee}}{1977}]{Cowie1977}
{Cowie} L.~L.,  {McKee} C.~F.,  1977, \mn@doi [\apj] {10.1086/154911}, \href
  {https://ui.adsabs.harvard.edu/abs/1977ApJ...211..135C} {211, 135}

\bibitem[\protect\citeauthoryear{{Dubois} \& {Commer{\c{c}}on}}{{Dubois} \&
  {Commer{\c{c}}on}}{2016}]{Rames2016}
{Dubois} Y.,  {Commer{\c{c}}on} B.,  2016, \mn@doi [\aap]
  {10.1051/0004-6361/201527126}, \href
  {https://ui.adsabs.harvard.edu/abs/2016A&A...585A.138D} {585, A138}

\bibitem[\protect\citeauthoryear{{El-Badry}, {Ostriker}, {Kim}, {Quataert}  \&
  {Weisz}}{{El-Badry} et~al.}{2019}]{Badry2019}
{El-Badry} K.,  {Ostriker} E.~C.,  {Kim} C.-G.,  {Quataert} E.,   {Weisz}
  D.~R.,  2019, \mn@doi [\mnras] {10.1093/mnras/stz2773}, \href
  {https://ui.adsabs.harvard.edu/abs/2019MNRAS.490.1961E} {490, 1961}

\bibitem[\protect\citeauthoryear{{Fick}}{{Fick}}{1855}]{fick}
{Fick} A.,  1855, \mn@doi [Annalen der Physik] {10.1002/andp.18551700105},
  \href {https://ui.adsabs.harvard.edu/abs/1855AnP...170...59F} {170, 59}

\bibitem[\protect\citeauthoryear{{Gombosi}, {Jokipii}, {Kota}, {Lorencz}  \&
  {Williams}}{{Gombosi} et~al.}{1993}]{Gombosi1993}
{Gombosi} T.~I.,  {Jokipii} J.~R.,  {Kota} J.,  {Lorencz} K.,   {Williams}
  L.~L.,  1993, \mn@doi [\apj] {10.1086/172209}, \href
  {https://ui.adsabs.harvard.edu/abs/1993ApJ...403..377G} {403, 377}

\bibitem[\protect\citeauthoryear{{Gottlieb} \& {Groth}}{{Gottlieb} \&
  {Groth}}{1988}]{Gottlieb1988}
{Gottlieb} J.~J.,  {Groth} C.~P.~T.,  1988, \mn@doi [Journal of Computational
  Physics] {10.1016/0021-9991(88)90059-9}, \href
  {https://ui.adsabs.harvard.edu/abs/1988JCoPh..78..437G} {78, 437}

\bibitem[\protect\citeauthoryear{{Gudiksen} \& {Nordlund}}{{Gudiksen} \&
  {Nordlund}}{2002}]{Boris2002}
{Gudiksen} B.~V.,  {Nordlund} {\r{A}}.,  2002, \mn@doi [\apjl]
  {10.1086/341600}, \href
  {https://ui.adsabs.harvard.edu/abs/2002ApJ...572L.113G} {572, L113}

\bibitem[\protect\citeauthoryear{{Hanasz} \& {Lesch}}{{Hanasz} \&
  {Lesch}}{2003}]{hanasz2003}
{Hanasz} M.,  {Lesch} H.,  2003, \mn@doi [\aap] {10.1051/0004-6361:20031433},
  \href {https://ui.adsabs.harvard.edu/abs/2003A&A...412..331H} {412, 331}

\bibitem[\protect\citeauthoryear{{Jubelgas}, {Springel}  \& {Dolag}}{{Jubelgas}
  et~al.}{2004}]{jubelgas2004}
{Jubelgas} M.,  {Springel} V.,   {Dolag} K.,  2004, \mn@doi [\mnras]
  {10.1111/j.1365-2966.2004.07801.x}, \href
  {https://ui.adsabs.harvard.edu/abs/2004MNRAS.351..423J} {351, 423}

\bibitem[\protect\citeauthoryear{Keller}{Keller}{2021}]{Pytipsy}
Keller B.,  2021, Pytipsy, \url{https://github.com/bwkeller/pytipsy}

\bibitem[\protect\citeauthoryear{{Keller}, {Wadsley}, {Benincasa}  \&
  {Couchman}}{{Keller} et~al.}{2014}]{Keller_2014}
{Keller} B.~W.,  {Wadsley} J.,  {Benincasa} S.~M.,   {Couchman} H.~M.~P.,
  2014, \mn@doi [\mnras] {10.1093/mnras/stu1058}, \href
  {https://ui.adsabs.harvard.edu/abs/2014MNRAS.442.3013K} {442, 3013}

\bibitem[\protect\citeauthoryear{{Kurganov} \& {Tadmor}}{{Kurganov} \&
  {Tadmor}}{2000}]{Kurg_Tad_2000}
{Kurganov} A.,  {Tadmor} E.,  2000, \mn@doi [Journal of Computational Physics]
  {10.1006/jcph.2000.6459}, \href
  {https://ui.adsabs.harvard.edu/abs/2000JCoPh.160..241K} {160, 241}

\bibitem[\protect\citeauthoryear{{Meyer}, {Balsara}  \& {Aslam}}{{Meyer}
  et~al.}{2012}]{Meyer2012}
{Meyer} C.~D.,  {Balsara} D.~S.,   {Aslam} T.~D.,  2012, \mn@doi [\mnras]
  {10.1111/j.1365-2966.2012.20744.x}, \href
  {https://ui.adsabs.harvard.edu/abs/2012MNRAS.422.2102M} {422, 2102}

\bibitem[\protect\citeauthoryear{{Monaghan}}{{Monaghan}}{1992}]{Monaghan1992}
{Monaghan} J.~J.,  1992, \mn@doi [\araa] {10.1146/annurev.aa.30.090192.002551},
  \href {https://ui.adsabs.harvard.edu/abs/1992ARA&A..30..543M} {30, 543}

\bibitem[\protect\citeauthoryear{{Monaghan}}{{Monaghan}}{1997}]{Monaghan1997}
{Monaghan} J.~J.,  1997, \mn@doi [Journal of Computational Physics]
  {10.1006/jcph.1997.5732}, \href
  {https://ui.adsabs.harvard.edu/abs/1997JCoPh.136..298M} {136, 298}

\bibitem[\protect\citeauthoryear{{Navarro}, {Khomenko}, {Modestov}  \&
  {Vitas}}{{Navarro} et~al.}{2022}]{Navarro_2022}
{Navarro} A.,  {Khomenko} E.,  {Modestov} M.,   {Vitas} N.,  2022, \mn@doi
  [\aap] {10.1051/0004-6361/202243439}, \href
  {https://ui.adsabs.harvard.edu/abs/2022A&A...663A..96N} {663, A96}

\bibitem[\protect\citeauthoryear{{Pakmor}, {Bauer}  \& {Springel}}{{Pakmor}
  et~al.}{2011}]{Arepo_2011}
{Pakmor} R.,  {Bauer} A.,   {Springel} V.,  2011, \mn@doi [\mnras]
  {10.1111/j.1365-2966.2011.19591.x}, \href
  {https://ui.adsabs.harvard.edu/abs/2011MNRAS.418.1392P} {418, 1392}

\bibitem[\protect\citeauthoryear{{Pontzen}, {Ro{\v{s}}kar}, {Stinson}  \&
  {Woods}}{{Pontzen} et~al.}{2013}]{pynbody}
{Pontzen} A.,  {Ro{\v{s}}kar} R.,  {Stinson} G.,   {Woods} R.,  2013, {pynbody:
  N-Body/SPH analysis for python}, Astrophysics Source Code Library, record
  ascl:1305.002 (\mn@eprint {ascl} {1305.002})

\bibitem[\protect\citeauthoryear{{Rempel}}{{Rempel}}{2017}]{Rempel2017}
{Rempel} M.,  2017, \mn@doi [\apj] {10.3847/1538-4357/834/1/10}, \href
  {https://ui.adsabs.harvard.edu/abs/2017ApJ...834...10R} {834, 10}

\bibitem[\protect\citeauthoryear{{Sedov}}{{Sedov}}{1959}]{sedov1959}
{Sedov} L.~I.,  1959, {Similarity and Dimensional Methods in Mechanics}

\bibitem[\protect\citeauthoryear{Snodin, Brandenburg, Mee  \& Shukurov}{Snodin
  et~al.}{2006}]{snodin2006}
Snodin A.~P.,  Brandenburg A.,  Mee A.~J.,   Shukurov A.,  2006, \mn@doi
  [MNRAS] {10.1111/j.1365-2966.2006.11034.x}, 373, 643

\bibitem[\protect\citeauthoryear{{Spitzer}}{{Spitzer}}{1956}]{Spitzer1956}
{Spitzer} L.,  1956, {Physics of Fully Ionized Gases}

\bibitem[\protect\citeauthoryear{{Taylor}}{{Taylor}}{1950}]{Taylor1950}
{Taylor} G.,  1950, \mn@doi [Proceedings of the Royal Society of London Series
  A] {10.1098/rspa.1950.0049}, \href
  {https://ui.adsabs.harvard.edu/abs/1950RSPSA.201..159T} {201, 159}

\bibitem[\protect\citeauthoryear{{Vishniac}}{{Vishniac}}{1983}]{vishniac1983}
{Vishniac} E.~T.,  1983, \mn@doi [\apj] {10.1086/161433}, \href
  {https://ui.adsabs.harvard.edu/abs/1983ApJ...274..152V} {274, 152}

\bibitem[\protect\citeauthoryear{{Wadsley}, {Stadel}  \& {Quinn}}{{Wadsley}
  et~al.}{2004}]{Wadsley2004}
{Wadsley} J.~W.,  {Stadel} J.,   {Quinn} T.,  2004, \mn@doi [\na]
  {10.1016/j.newast.2003.08.004}, \href
  {https://ui.adsabs.harvard.edu/abs/2004NewA....9..137W} {9, 137}

\bibitem[\protect\citeauthoryear{{Wadsley}, {Keller}  \& {Quinn}}{{Wadsley}
  et~al.}{2017}]{Wadsley_2017}
{Wadsley} J.~W.,  {Keller} B.~W.,   {Quinn} T.~R.,  2017, \mn@doi [\mnras]
  {10.1093/mnras/stx1643}, \href
  {https://ui.adsabs.harvard.edu/abs/2017MNRAS.471.2357W} {471, 2357}

\bibitem[\protect\citeauthoryear{{Weaver}, {McCray}, {Castor}, {Shapiro}  \&
  {Moore}}{{Weaver} et~al.}{1977}]{Weaver1977}
{Weaver} R.,  {McCray} R.,  {Castor} J.,  {Shapiro} P.,   {Moore} R.,  1977,
  \mn@doi [\apj] {10.1086/155692}, \href
  {https://ui.adsabs.harvard.edu/abs/1977ApJ...218..377W} {218, 377}

\bibitem[\protect\citeauthoryear{{Zamora}, {Slaughter}  \& {Abel}}{{Zamora}
  et~al.}{2023}]{zamora2023}
{Zamora} A.,  {Slaughter} E.,   {Abel} T.,  2023, \mn@doi [\mnras]
  {10.1093/mnras/stad770}, \href
  {https://ui.adsabs.harvard.edu/abs/2023MNRAS.521.3186Z} {521, 3186}

\makeatother
\end{thebibliography}



\appendix

\counterwithin{figure}{section}

\section{Exact Solutions} \label{appendix_a}
In this paper, we make use of the exact solution for hyperbolic conduction. This is done for the case where $\tau$, $\rho$, and $\kappa$ are all kept constant. From separation of variables, we can change equation~\ref{eq:hyp_cond_full} from the main text into the following system of equations.

\begin{equation}
    \begin{aligned}
    u(x,t) = X(x)T(t)
    \end{aligned}
    \label{eq:sep_var_1}
\end{equation}
\begin{equation}
    \begin{aligned}
    \frac{1}{T}\frac{d^2T}{dt^2}+\frac{1}{\tau T}\frac{dT}{dt} = -\alpha^2
    \end{aligned}
    \label{eq:sep_var_1}
\end{equation}
\begin{equation}
    \begin{aligned}
    \frac{\kappa}{\rho\tau X}\nabla^2 X = -\alpha^2
    \end{aligned}
    \label{eq:sep_var_1}
\end{equation}
Where $-\alpha^2$ is some constant of integration. From this we can find that the general solution to equation~\ref{eq:hyp_cond_full} is:
\begin{equation}
    \begin{aligned}
    u = u_0 + \sum_{n=1}^\infty\left[A_n\cos{\left(\frac{2\pi n}{L}x\right)} +
    B_n\sin{\left(\frac{2\pi n}{L}x\right)}\right] 
    \\
    \times \left[\cosh{\left(\frac{\gamma_n}{2\tau}t\right)}+\frac{1}{\gamma_n}\sinh{\left(\frac{\gamma_n}{2\tau}t\right)}\right]e^{-\frac{t}{2\tau}}
    \end{aligned}
    \label{eq:hyp_cond_exact}
\end{equation}
\begin{equation}
    \gamma_n^2 = 1 - 16\left(\frac{\pi n}{L}\right)^2\frac{\kappa\tau}{\rho}
    \label{eq:gamma_cond}
\end{equation}
Where $n$ is the wave number of each member of the Fourier series. 
Note that $\gamma_n^2 \leq 0$ is an admissible answer because, in cases where $\gamma_n$ is either 0 or is purely imaginary, u still has a purely real answer. 

Using a similar process for the parabolic system nets us the following exact solution:
\begin{equation}
    \begin{aligned}
    u = u_0 + \sum_{n=1}^\infty \left[A_n \cos\left(\frac{2\pi n}{L}x\right)+B_n\sin\left(\frac{2\pi n}{L}x\right)\right]e^{-\left(\frac{2\pi n}{L}\right)^2\frac{\kappa}{\rho} t}
    \end{aligned}
    \label{eq:par_cond_exact}
\end{equation}

\section{Derivation of Signal Speed}\label{appendix_b}

We make the assumption, for each particle, the time step will be small enough that $\rho$, $\tau$, and $\kappa$ will not change significantly in a short period of time and can therefore be treated as constant. 
Of course, this is not always true, but does still work remarkably well in practice.

For a hyperbolic equation, the signal speed, c, is the minimum eigenvalue of the Jacobian matrix.
To determine this, we can look at a rearranged version of our equations~\ref{eq:par_cond_2} and~\ref{eq:hyp_cond} in one dimension.
\begin{align}
    \frac{\partial}{\partial t}\begin{bmatrix}
           u \\
           Q\\
         \end{bmatrix}
         + \frac{\partial}{\partial x} \begin{bmatrix}
           \frac{1}{\rho}Q\\
           \frac{\kappa}{\tau} u\\
         \end{bmatrix}
         = \begin{bmatrix}
           0 \\
           \frac{1}{\tau}Q\\
         \end{bmatrix}
         \label{eq:cons_equ}
  \end{align}

Note that, in this case, we assume that on short scales, $\tau$, $\kappa$, and density are constant. 
The Jacobian matrix corresponding to this is:
\begin{align}
    J = \begin{bmatrix}
        0 \ \frac{\kappa}{\tau}\\
        \frac{1}{\rho} \ 0\\
    \end{bmatrix}
    \label{eq:jacobian}
\end{align}
Which has eigen-values of $\pm \sqrt{\frac{\kappa}{\rho\tau}}$. 
Thus, signal speed is $\sqrt{\frac{\rho \tau}{\kappa}}$.




\bsp	
\label{lastpage}
\end{document}